\documentclass{IEEEtran}
\usepackage{cite}
\usepackage{amsmath,amssymb,amsfonts}
\usepackage{graphicx}
\usepackage{textcomp,nicefrac}
\def\BibTeX{{\rm B\kern-.05em{\sc i\kern-.025em b}\kern-.08em
T\kern-.1667em\lower.7ex\hbox{E}\kern-.125emX}}
\begin{document}
\title{The GosipGUI framework for control and benchmarking of readout 
electronics front-ends} 
\author{J\"orn Adamczewski-Musch and Nikolaus Kurz
\thanks{J\"orn Adamczewski-Musch and Nikolaus Kurz are with the Experiment Electronics Department, GSI Helmholtzzentrum f. Schwerionenforschung GmbH, Planckstr. 1, D-64291 Darmstadt, Germany. The corresponding author is J\"orn Adamczewski-Musch (e-mail: j.adamczewski@gsi.de)}
}
\maketitle

\begin{abstract}
The GOSIP (Gigabit Optical Serial Interface Protocol) provides communication via optical fibres between multiple kinds of front-end electronics and the KINPEX PCIe receiver board located in the readout host PC. In recent years a stack of device driver software has been developed to utilize this hardware for several scenarios of data acquisition. On top of this driver foundation, several graphical user interfaces (GUIs) have been created.
These GUIs are based on the Qt graphics libraries and are designed in a modular way: All common functionalities, like generic I/O with the front-ends, handling of configuration files, and window settings, are treated by a framework class GosipGUI. In the Qt workspace of such GosipGUI frame, specific sub classes may implement additional windows dedicated to operate different GOSIP front-end modules. These readout modules developed by GSI Experiment Electronics department are for instance FEBEX sampling ADCs, TAMEX FPGA-TDCs, or POLAND QFWs.
For each kind of front-end the GUIs allow to monitor specific register contents, to set up the working configuration, and to interactively change parameters like sampling thresholds during data acquisition. The latter is extremely useful when qualifying and tuning the front-ends in the electronics lab or detector cave.
Moreover, some of these GosipGUI implementations have been equipped with features for mostly automatic testing of ASICs in a prototype mass production. This has been applied for the APFEL-ASIC component of the PANDA experiment currently under construction, and for the FAIR beam diagnostic readout system POLAND.

\end{abstract}


\section{Introduction}
\label{sec:intro}
\begin{figure}[htb]
\centerline{\includegraphics[width=1.1\columnwidth]{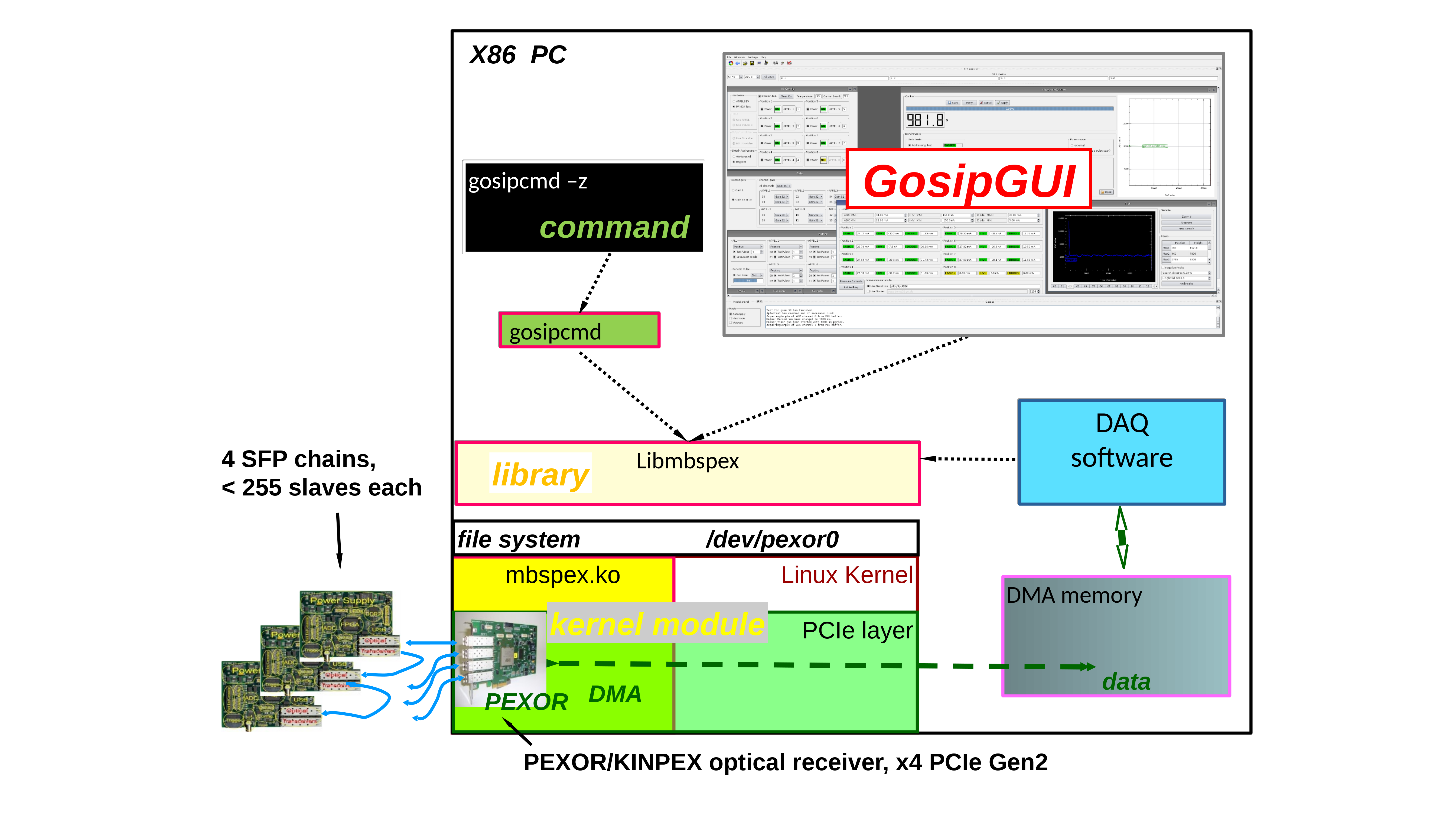}}
\caption{Schematic view of software elements used for GOSIP communication. Summary picture taken from \cite{b-mbspex}. See text for details..}
\label{f-driversoftware}
\end{figure}

\IEEEPARstart{O}{ne} kind of GSI standard DAQ hardware system is set up with a PC based PCIe optical receiver board KINPEX that connects up to 4 optical fibre chains of up to 256 readout slave boards each,  digitizing the detector data. Communication between the KINPEX and the front-end devices is handled with the protocol GOSIP (Gigabit Optical Serial Interface Protocol) \cite{b-gosip}. 
Linux device driver and higher level software \cite{b-mbspex} has been provided for different use cases of data acquisition, both for experiment detector readout with frameworks like MBS \cite{b-mbs}, and beam diagnostic systems with FESA \cite{b-fesa}. This software consists of a kernel module of the KINPEX board, user libraries in C and C++, the command line tool \emph{gosipcmd} for front-end register access, and graphical user interfaces (GUIs) tailored specifically for the front-end hardware connected to the optical fibre chains.  
Fig. \ref{f-driversoftware} gives a schematic overview of these software elements.

These GUIs are meant to interactively monitor, control and configure various kinds of GOSIP slave devices. 
Most recently the collection of different GUI software has been redesigned to fit into a common {\em GosipGUI} framework.

\section{Software architecture}

\subsection{Base environment}
The Qt graphics framework \cite{b-qtframework} was initially chosen for implentation of several required GUIs controlling the GOSIP hardware, since it allows rapid development by means of the {\em designer} tool. Moreover, it is freely available on most Linux platforms, and the C and C++ APIs of the mbspex and pexor driver libraries can be easily called in the Qt slot functions.

\subsection{Object collaborations}
\label{subsec:objects}
\begin{figure}[htb]
\centerline{\includegraphics[width=1.2\columnwidth]{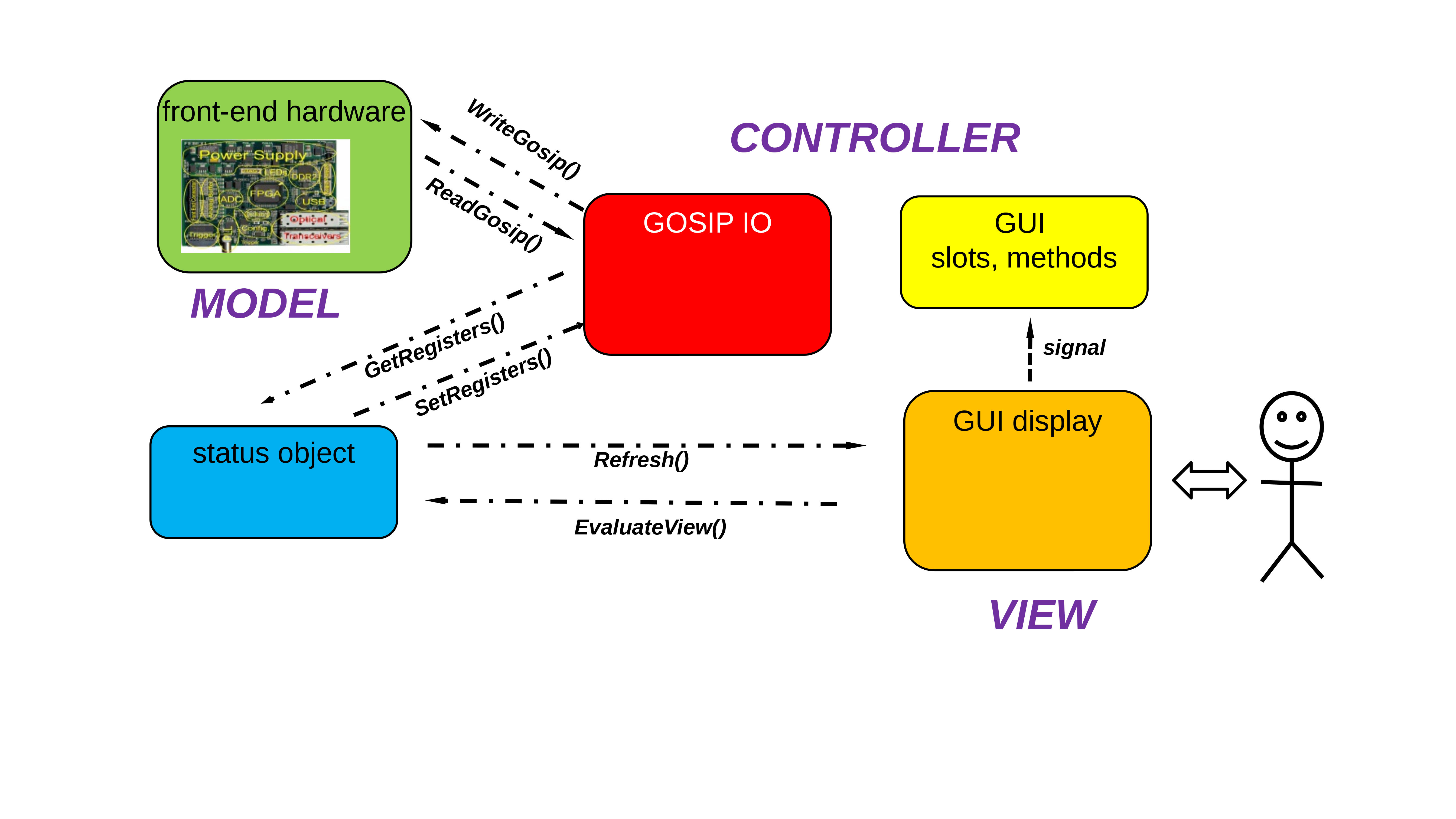}}
\caption{Collaboration diagram of GosipGUI objects. See text for details}
\label{f-collaboration}
\end{figure}

Figure \ref{f-collaboration} shows a collaboration diagram of the participating entities of the {\em GosipGUI} software: Communication with the \emph{front-end-hardware} under control is handled by a functional {\em GOSIP IO} block, implementing fundamental methods {\em ReadGosip()} and {\em WriteGosip()}. This part is using either direct calls of driver library functions, or may for certain use cases invoke 
{\em gosipcmd} scripts (see Fig.\ref{f-driversoftware}). The {\em GOSIP IO} block is the same for all front-end implementations and part of the framework base classes.

The role of the \emph{status object} is to buffer register contents and state of the controlled \emph{front-end-hardware} in the GUI process memory. It may also aggregate additional objects for characterisation of the device under control, like measurement results. Therefore the \emph{status object} is a specific implementation for the kind of GOSIP slave. For each slave connected to the optical fibre, there is one instance of this \emph{status object}. The framework handles all existing \emph{status object}s by a generic collection of base class pointers. 

The \emph{GUI display} can visualize the state of only one front-end slave at once. When the human operator chooses the slave by SFP and device number on the GUI, the corresponding \emph{status object} is selected, updated from the hardware by means of the \emph{GetRegisters()} method, and the GUI contents are refreshed from the \emph{status object} by method \emph{Refresh()}. When the user changes values on the GUI interactively, firstly the \emph{status object} is modified by means of Qt ``signal-slot'' callbacks, or by using the global method \emph{EvaluateView()}. In regular GUI mode, all changes done by the operator are applied to the hardware slave by \emph{SetRegisters()} function not before the user explicitely presses an ``Apply'' button. However, in a special ``AutoApply'' mode this buffering mechanism can be overruled and single register values are written immediately to the hardware when they are changed on the GUI. Still the \emph{status object} is always consistent with the hardware state also in this mode.

Although the {\em GosipGUI} can visualize one selected slave only, it is possible to change all connected devices to the same register values in a kind of ``multicast'' or ``broadcast mode''. This means that the visible contents of the GUI are first copied to all \emph{status object}s under the multicast addressing scheme, and then immedieately to the \emph{front-end-hardware} devices. Such multicast may cover all slaves at a selected SFP chain, or all slaves at all chains. Moreover, also the ``AutoApply'' mode can be invoked with multicast addressing.

Besides this, the registry of \emph{status object} allows to save the state of all connected slaves to a setup script file. Instead of writing the contents of the \emph{status object}s to the hardware using \emph{SetRegisters()}, a similar function \emph{SaveRegisters()} will translate the status values into a series of \emph{gosipcmd} commands that is dumped to a \texttt{*.gos} file. This command file can be invoked to recover the register values, either from a plain shell with \texttt{gosipcmd -f file.gos}, or from within the {\em GosipGUI} environment.

The relations between the setup object, the specific front-end GUI elements, and the generic \emph{GosipGUI} buttons can be regarded as a kind of ''Model-View-Controller`` design pattern \cite{b-designpatterns}. This pattern is extended in a way that the ''model`` of the displayed data is not represented by \emph{front-end hardware} registers only, but it is buffered in GUI process memory by the \emph{status object}. The ''controller'' part that visualizes the register ``model'' in the GUI ``view'' consists of both generic components such as the {\em GOSIP IO} entity, and front-end specific functionalities that are implemented into the framework by means of virtual methods. 
The current \emph{GosipGUI} concept always uses that the same ``view'' (GUI implementation) for different ``model'' contents (i.~e. the selected hardware slave).
Moreover, depending on the application the GUI may also offer different ``views'' for the same ``model'' (slave register data), e.~g. a plain numerical display or a graphical plot.

\subsection{Class design}
\label{subsec:classes}
When the number of GOSIP slave device variants increased that had to be implemented as special GUIs, it turned out that many functionalities are naturally in common between them. Because of this, a first redesign of the originally independent Qt GUIs have been done such that all common I/O functionalities via the driver libs were moved into a framework base class {\tt  GosipGui}. This base class implements all functionalities of the generic {\em GOSIP IO} block (compare Fig. \ref{f-collaboration} in section \ref{subsec:objects}). It also provides all common GUI elements, as described in section \ref{sec:commongui}.

\begin{figure}[htb]
\centerline{\includegraphics[width=1.1\columnwidth]{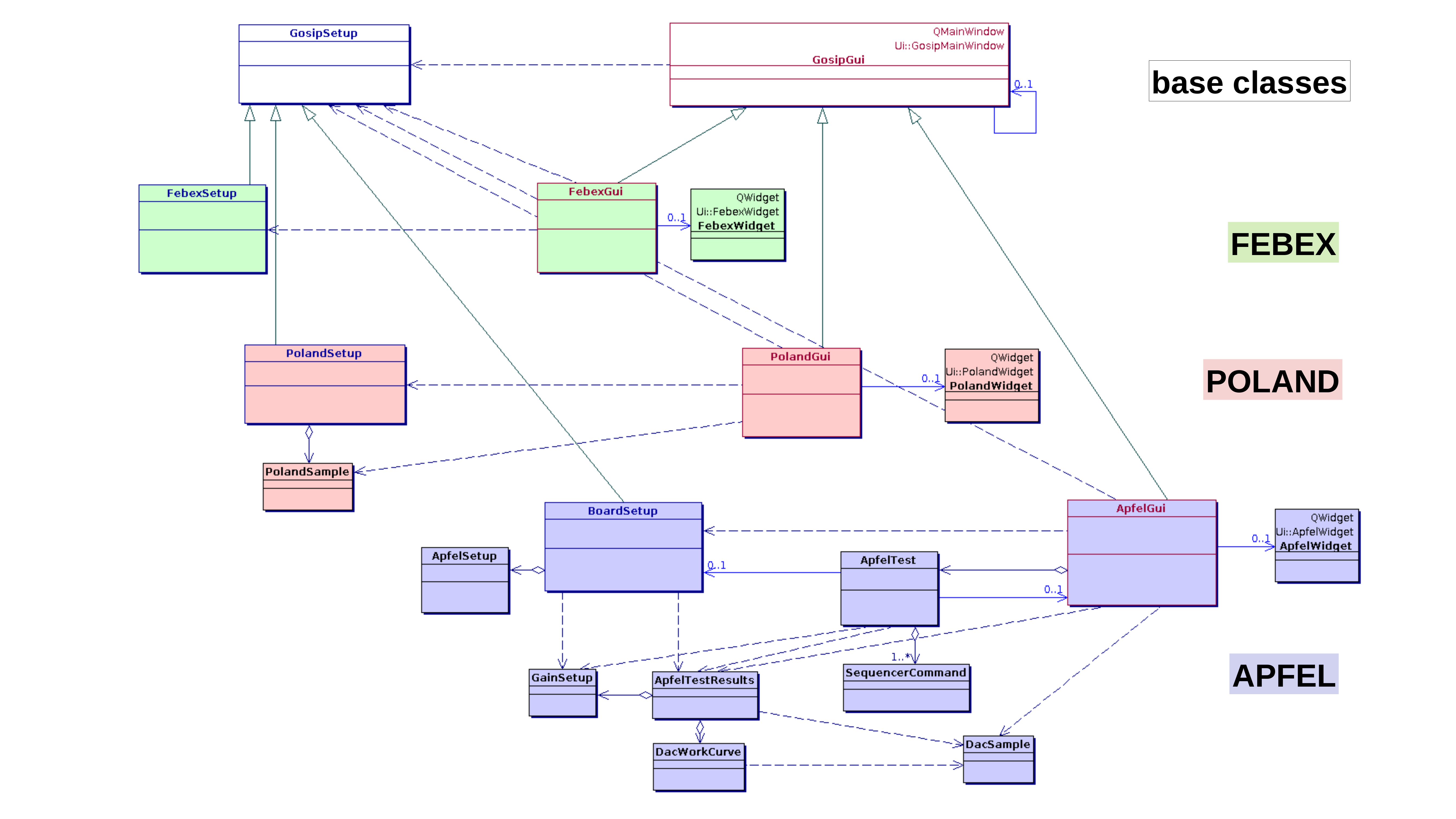}}
\caption{Class diagram of GosipGUI: 
The main base classes GosipSetup and GosipGui are implemented for different applications FEBEX, POLAND, and APFEL. See text for details.}
\label{f-classdiagram}
\end{figure}

\begin{figure*}[htb]
\centerline{\includegraphics[width=1.1\textwidth]{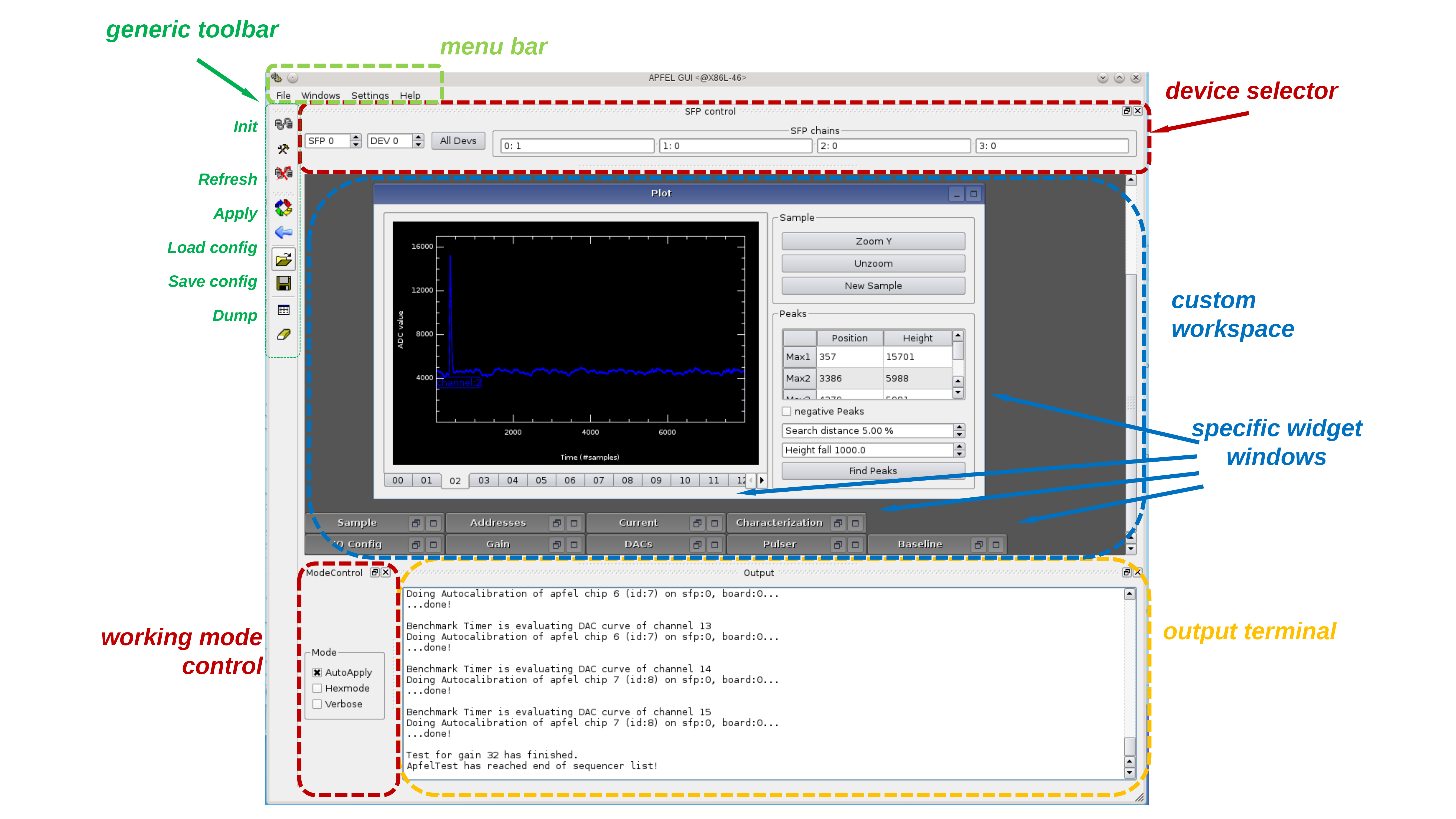}}
\caption{The common elements of the base class \emph{GosipGui} provide a frame for the custom workspace with subclass specific windows: The device selector sets the GOSIP slave device that is currently under control. The chosen working modes, like ``AutoApply'',  may rule the generic behaviour of GUI functionalities, and may modify the output verbosity on the text terminal. The menus and generic toolbar buttons will call virtual methods that can be modified in the implementations. See text for further details.}
\label{f-commongui}
\end{figure*}

Fig. \ref{f-classdiagram} contains a class diagram of the  \emph{GosipGUI} framework and some implementation examples. 
All special GUI implementations will inherit from this \texttt{GosipGui} base class in C++ fashion. This means, they can add further GUI elements that are required for their hardware regsisters only to this main window. These additional elements are composed in separate Qt designer projects for the subclass. They may either define an arbitrary number of \texttt{QTabWidget} entries in a reserved space of the base class \texttt{QMainWindow}, or may add special windows into the \texttt{QWorkspace} if the base class GUI is run in this mode. The latter ``desktop-like`` mode has been added recently with a second redesign of the \emph{GosipGUI} framework, additionally featuring \texttt{QSettings} to save and restore the internal window positions, and other Qt workspace functionality like iconifying, tiling and cascading all contained user windows.

Besides the visible extensions of the GUI elements, the subclasses of \texttt{GosipGui} can overwrite several virtual methods to redefine the functionalities of the generic framework buttons,  for example {\em Apply()} (send setup shown in GUI to the selected gosip slave), {\em Refresh()} (retrieve register contents of slave and show current state on GUI), and {\em Dump()} (Printout of any reasonable registers to the logging console).

Moreover, the framework also defines a base class {\tt GosipSetup} that represents the \emph{status object} described in section \ref{subsec:objects}. 
The {\tt  GosipGui} base class manages these \emph{status object}s for all existing slaves in a collection of pointers to {\tt GosipSetup} instances.
Moreover, it provides a factory method \cite{b-designpatterns} for {\tt GosipSetup}
that has to be implemented in the GUI subclass for the appropriate hardware representation. 

As seen in Fig.\ref{f-classdiagram}, the GUI implementations (here for FEBEX, POLAND, and APFEL front-end hardware) may not only define subclasses of \texttt{GosipGui} and {\tt GosipSetup}. Additionally, they may also add proprietary classes for special purposes, e.~g. aggregating subcomponents of the slave registers, or recording temporary data acquisition samples that are not strictly register contents.

\begin{figure*}[htb]
\centerline{\includegraphics[width=0.95\textwidth]{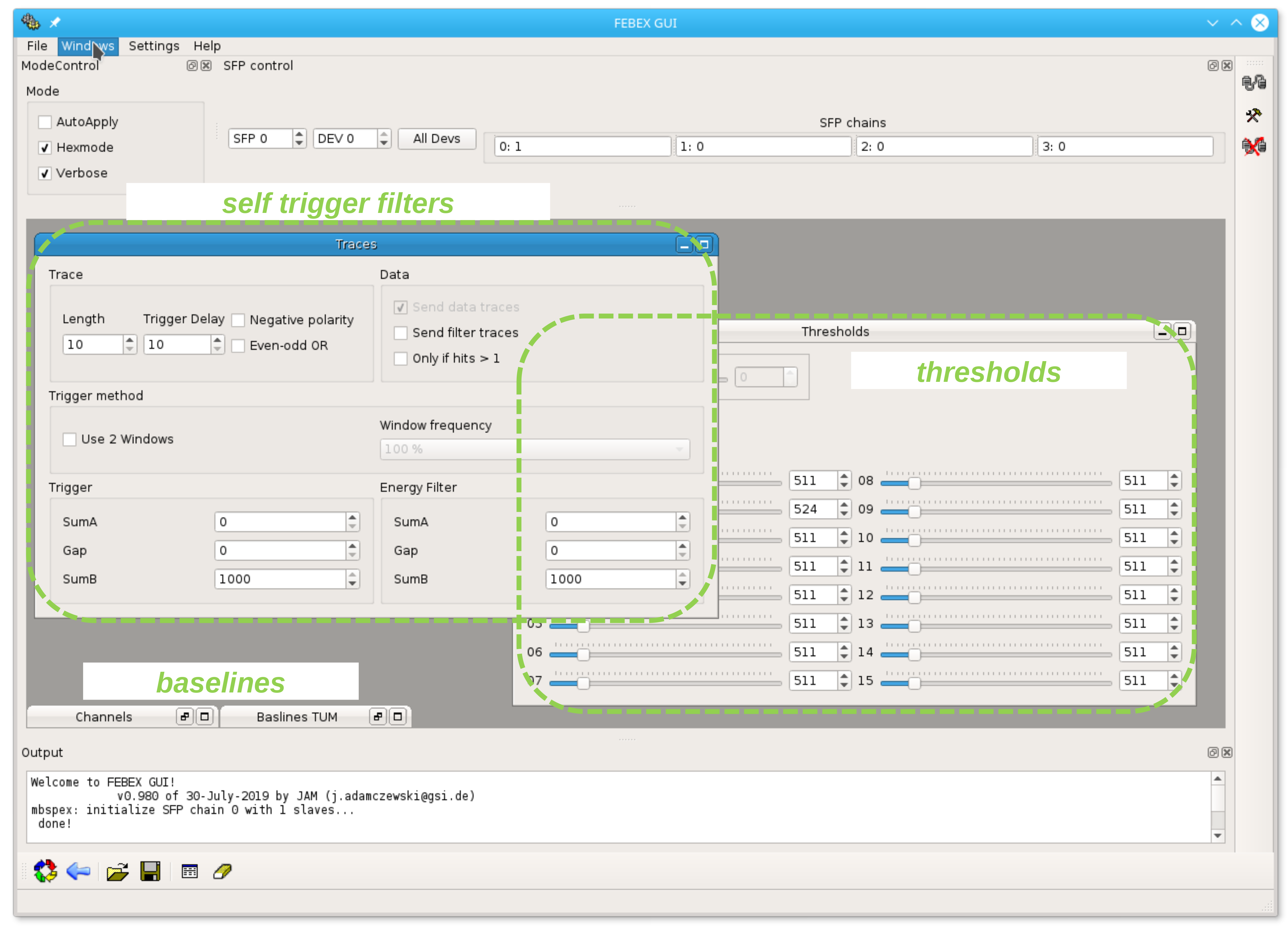}}
\caption{The FebexGui in workspace mode.  Several special windows allow to control the hit filter, thresholds, and the baselines of a DAC addon board. See text for details}
\label{f-febexgui}
\end{figure*}

\begin{figure*}[htb]
\centerline{\includegraphics[width=0.95\textwidth]{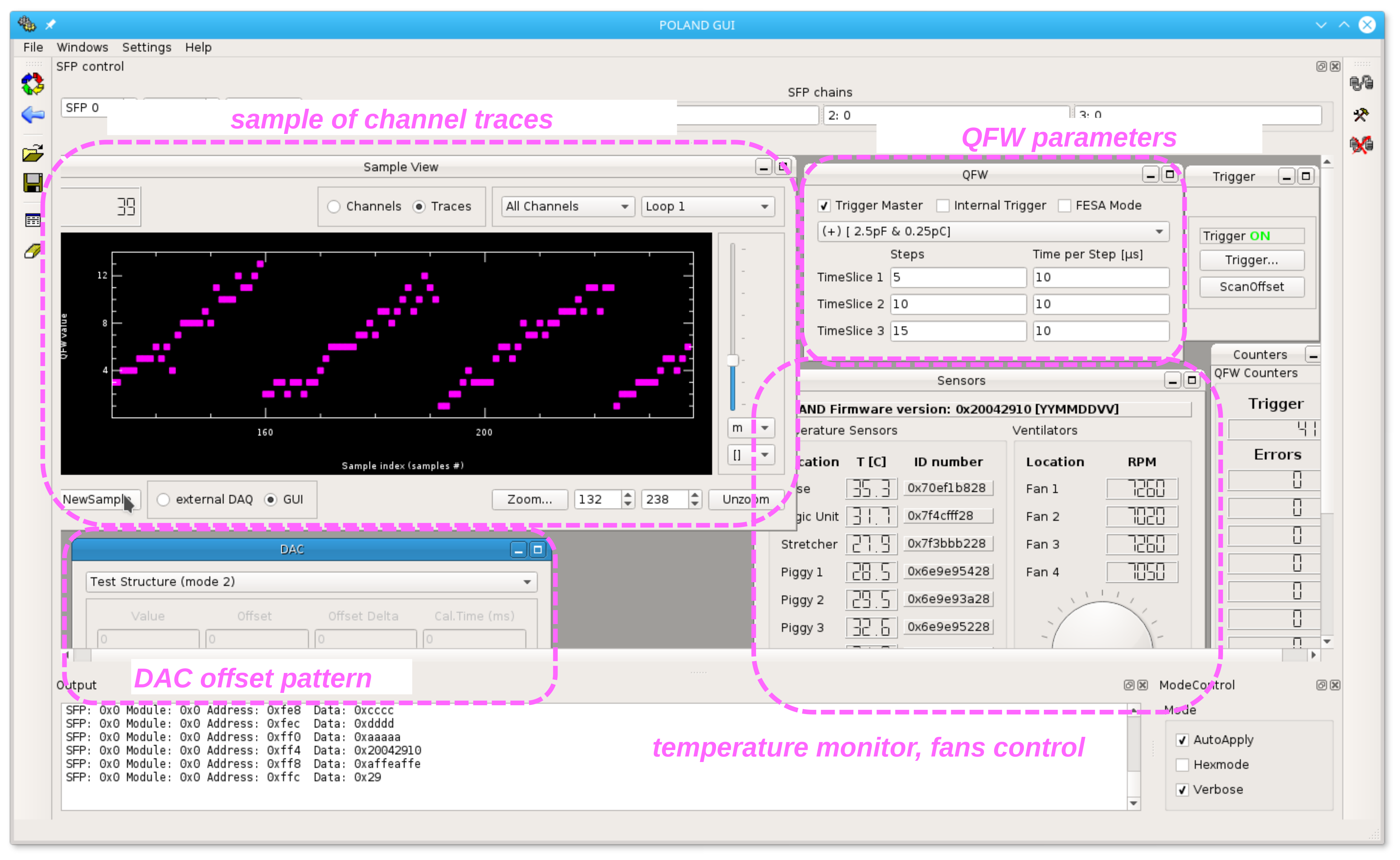}}
\caption{The PolandGui in workspace mode with several special windows: The QFW parameters, like time slice duration and charge sensitivity, can be set. The DAC offset values can be specified for each channel and may produce a steplike test pattern. A sample of the channel traces can be acquired and visualized. Several on-board sensors and fans can be monitored. See text for further details.}
\label{f-polandgui}
\end{figure*}

\subsection{Coding macros}
For developer's convenience, the {\em GosipGUI} framework offers several C preprocessor macros that are helpful to define repeated functionality for different subclasses and functions. For example, retrieving and checking the currently active {\tt GosipSetup} object from the \texttt{GosipGui} slave registry and downcasting it to the correct subclass requires always the same code snippets in all functions of the \texttt{GosipGui} subclass that want to use their status object. Another example is the invocation of any subclass member function after checking some state or condition of the base class, or the repeated invocation of any member function for all connected slaves (''broadcast''). The following list gives a brief overview of the implemented macros:

\begin{itemize}

 \item \texttt{theSetup\_GET\_FOR\_SLAVE(X)} :  get currently active \texttt{GosipSetup} object from base class and cast it to actual implementation class \emph{X}. This provides a local handle \texttt{theSetup->} in the invoking function to access any special member of \emph{X}
 
 \item \texttt{GOSIP\_BROADCAST\_ACTION(X)} : execute function \emph{X} for the currently selected front-end. The macro will evaluate the active slave selection of the common GUI and can optionally perform this function in a ``multicast'' mode for all devices of an SFP chain, or a ``broadcast''for all connected devices at all chains. Although the pexor kernel module already implements a single write multicast functionality that can be used from plain \emph{gosipcmd}, it turned out that this is not sufficient when more complex operations, like well defined write sequences, shall be carried out for all slaves
 
 \item \texttt{GOSIP\_AUTOAPPLY(X)}: execute function \emph{X} for the currently selected front-end only if the common GUI is in the “auto-apply” mode. Used for interactive tuning of single registers without writing the complete setup to hardware. This macro implies the above \texttt{GOSIP\_BROADCAST\_ACTION(X)}.
 
\end{itemize}

It should be mentioned that the functionalities of such C preprocessor macros might as well be implemented by means of certain C++ templates in the base class. However, during the practical development work for new {\tt  GosipGui} subclasses there is no difference between writing a macro expression or defining a template function for a new type argument. So the initially chosen macro technique has not been improved yet.

\section{Common Gui elements}
\label{sec:commongui}
The {\tt  GosipGui} base class, inherited from {\tt QMainWindow}, contains all GUI elements used by all subclasses, e.~g. the spinboxes defining the currently active SFP chain and front-end slave, tools for reset and initialization of GOSIP communication, and buttons for apply and refresh between the GUI display and the controlled hardware. Additionally, a status message line and a logging console for generic printouts are embedded into the base class GUI frame. Last but no least the base class defines several ``working modes'' of the GUI, for example the displayed number representation (decimal or hex), the debug output of I/O operations to the log window, or the so-called ``auto apply'' mode. This mode will let all changes on the GUI be written to the slave hardware immediately, which is useful for interactive tweaking of single register values without the need to transfer the complete setup. In Fig. \ref{f-commongui} a screenshot of these common elements provided by the {\tt  GosipGui} base class is shown.

\section{Application examples}
For almost each type of GOSIP front-end slaves there is currently an implementation of \emph{GosipGUI}. This covers experiment DAQ readout boards, like the FEBEX sampling ADC (section \ref{sub_febex}), or the POLAND QFW charge-frequency converter (section \ref{sub_poland}).
Initially  meant for inspecting a few boards in a laboratory or small experiment setup only, some \emph{GosipGUI} implementations have been extended for much more complex tasks. As an example,
section \ref{sub_apfel} treats the application for characterization of the APFEL ASIC mass production.

\subsection{FEBEX}
\label{sub_febex}
The FEBEX family of DAQ boards host a 12 or 14-bit pipe-lining ADC with 16 channels and up to 100~MHz sampling frequency. An FPGA (Lattice LFE3 for FEBEX3, or Xilinx FPGA XC7 for FEBEX4) implements the GOSIP data transfer and provides a programmable fast trapezoidal filters for hit finding, and energy trapezoidal filter for energy measurement of each detected hit \cite{b-febex}. So FEBEX can be applied for data acquisition with a local ``self-triggered`` readout, and provide a feature extraction for data reduction. Different addon-boards allow to equip FEBEX for a large number of experimental use cases. 

Figure \ref{f-febexgui} displays a screenshot of the \emph{FebexGui} with special control windows for basic FEBEX functionalities, like setting channelwise thresholds, or defining the hit filter parameters. It may also control the base line level for each input channel, provided by a special addon-board with DACs. The \emph{FebexGui} implements an iterative algorithm to set these DAC outputs to a desired value for each ADC input channel: While tuning the DACs, the ADC values are continuously measured until the difference to the nominal set point is below a predefined tolerance. It is also possible to tune all 16 channels to the same baseline value automatically.

\subsection{POLAND}
\label{sub_poland}
The POLAND hardware are intended for FAIR beam diagnostic systems using SEM grids \cite{b-poland}. These front-end boards contain ASICs with charge-frequency converters (QFW) for 32 input channels. Each channel can acquire three subsequent time slice samples with arbitrary number of steps, and a time granularity down to 1~$\mu$s.  Corresponding DACs can provide each channel with a charge bias for offset correction, and may be used to generate an internal test pattern.

\begin{figure*}[htb]
\centerline{\includegraphics[width=0.95\textwidth]{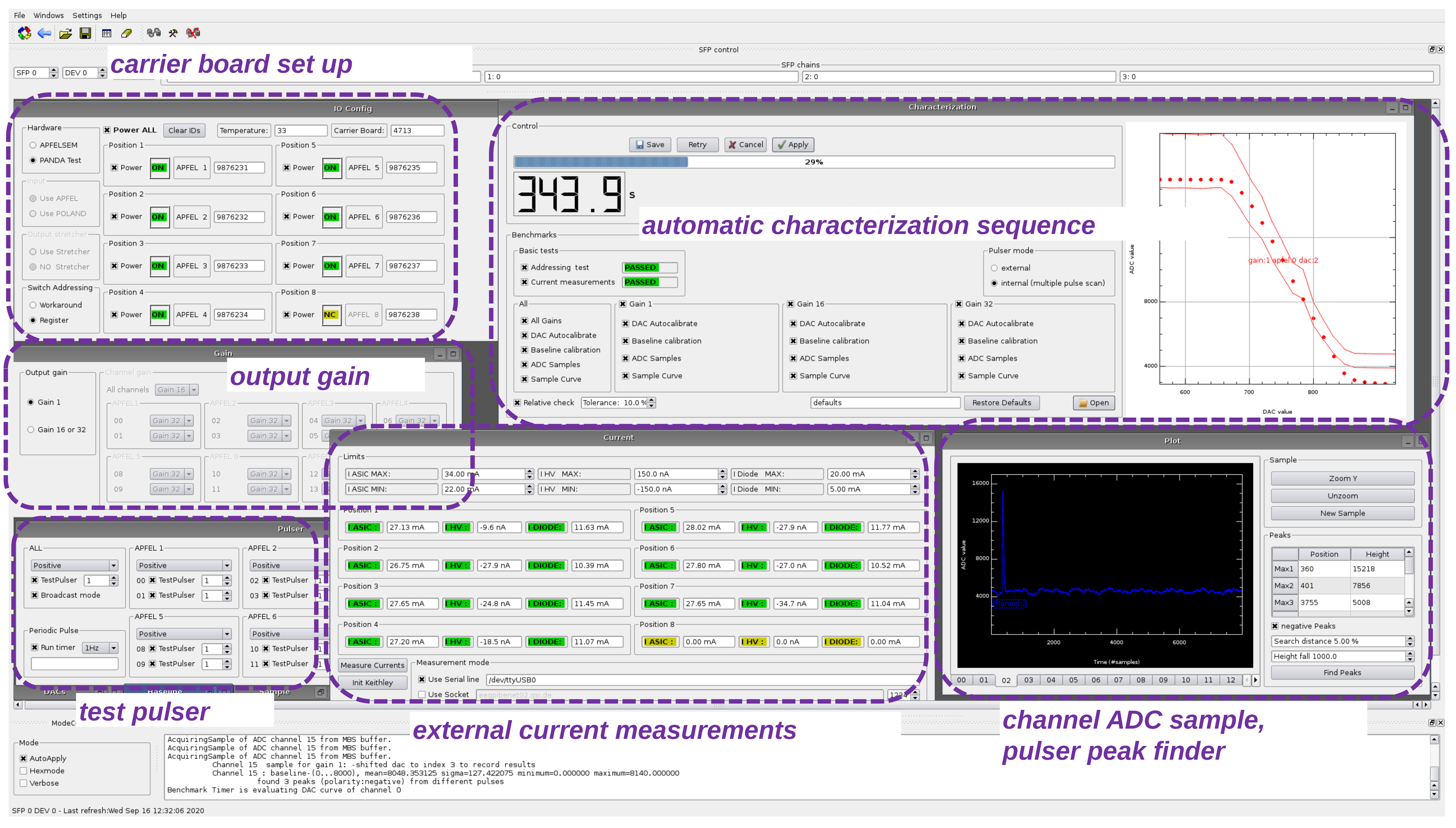}}
\caption{The ApfelGui in workspace mode with several special windows. The connected chips under test on the carrier board can be registered by scanning a QR-code id. Different gain modes or test pulser characteristics can be set manually. Samples of the ADC from MBS DAQ can be evaluated directly to draw figures of merit. The complete automatic characterization sequence of all aspects can be controlled by a central benchmarking window. See text for further details.}
\label{f-apfelgui}
\end{figure*}

Fig. \ref{f-polandgui} gives an impression of the most recent version of the \emph{PolandGui}. 
The frame of the main window contains dock windows with generic buttons and menus for the invocation of the special apply and refresh methods, executing \emph{gosipcommand} configuration scripts from file system, or defining the operation mode. The logging console and the status message at the bottom are also visible. Inside the workspace the tool windows dedicated for POLAND control can be arranged arbitrarily. These handle the setup of the sampling tracelength of the charge frequency converter, the DAC values  for the channel offset tests, and the display of temperatures and fan speeds. 

Most recently a functionality has been added to acquire a single trace sample from the QFW and plot it in a \emph{KPlotWidget} canvas. This sample plot window offers several display options, like sorting and accumulating the data of different channels and trace regions, and zooming and panning of the visible region by mouse interactions. The sample readout is carried out either from an external data acquisition system like MBS \cite{b-mbs} that runs in parallel, and \emph{PolandGui} just snoops the contents of the front-end readout buffers. Or another mode is possible, where  \emph{PolandGui} invokes the PEXOR driver library to fetch the current trace sample without the need of an external DAQ system. This direct DAQ functionality is  going to be used for mass testing of the already produced POLAND devices and may be extended by semi-automatical test procedures, as they are described in the following section \ref{sub_apfel}

\subsection {APFEL}
\label{sub_apfel}
The APFEL (ASIC for Panda Front-end ELectronics) 
offers an integrated charge sensitive preamplifier and shaper. It is optimized for the readout of avalanche photo diodes with large detector capacitance and high event rates \cite{b-holgerpaper}.
The \emph{ApfelGui} is dedicated for test carrier boards of this APFEL, using a FEBEX based system for data acquisition (see section \ref{sub_febex}).

Fig.\ref{f-apfelgui} shows a screenshot of the \emph{ApfelGui} that is capable of production quality benchmarking of ASICs in mass production.
The GUI offers semi-automatical test procedures for simultaneous characterization of 8 chips, including evaluation of both digital and analog data \cite{b-apfel}. This consists in basic functionality tests of registers, current measurements with external devices, evaluation of DAC working curves, and sampling ADC readout of pulser generated shapes by accessing the buffers of a concurrently running MBS DAQ system \cite{b-mbs}. For this purpose, a custom command sequencer has been implemented that can invoke the (usually interactive) GUI actions required for the benchmark from within a Qt timer callback. 
Before starting the sequence, the human operator will scan the IDs of the chips under test from their QR-code, and may chose the kind of performed checks on the GUI. During the tests the results of the various benchmarks are immediately visualized on the GUI. If some values are outside their predefined nominal range, this will show up at once as ''red lamp``, or as working curve outside a range corridor in the plotted canvas. 
After finishing the characterization sequence, the figures of merit can be stored together with the chip IDs as ASCII formatted file for later database export.  
Typically, the complete characterization of 8 ASICs takes about 14 minutes.
Since 2017 this system has been used for commissioning of about 2000 APFEL ASICs to be delivered for the PANDA experiment at FAIR.

\section{Conclusions and Outlook}
The \emph{GosipGUI} framework has been developed as a common graphical user interface system for various front-ends using the GOSIP data transfer protocol. The partitioning between generic base class functionality and specific front-end subclasses has proved to be a powerful design to minimize the effort of implementing GUIs for new kinds of boards with the same communication layer.

The \emph{GosipGUI} concept to visualize and control just one slave device at once was stemming from the requirements of small test setups in an electronics lab. Here often a single device only is connected to the SFP chain and should be tuned and tested interactively by the developers. In this environment such GUI is very helpful for checking the properties of the boards, especially when a concurrent DAQ system with online monitoring provides live data samples during change of readout board parameters. Besides the manual operator intervention, the \emph{GosipGUI} may implement even automatic test sequences for large number of devices in a deployment phase, as the APFEL example (section \ref{sub_apfel}) demonstrates. This shows the capabilities of the GOSIP software in connection with the Qt5 development environment. Although this has been very useful for the described application, it probably can not compete with other frameworks especially tailored for such automatic test procedures in a large scale industrial scope.

Concerning real experiment configurations in an accelarator facility cave, still the \emph{GosipGUI} can be useful to check single front-end entities of a larger set-up. The user may also tune the general parameters, like channel thresholds, with the \emph{GosipGUI} at one represantative front-end unit, and propagate the same values to all boards by means of the \emph{GosipGUI} broadcast features. On the other hand, each single board may as well be tuned individually in the setup phase of an experiment, and such values can be stored to the {\tt *.gos} setup script to apply them as automatic default at each DAQ startup.

However, in comparison with ''fully-grown`` control systems, e.~g. EPICS \cite{b-epics}, the current state of \emph{GosipGUI} has some disadvantages. At first, the \emph{GosipGUI} is capable of controlling front-ends only if they are locally connected to the PEXOR SFPs at the same host PC where it is running. Large distributed DAQ set-ups with many subevent readout PC nodes would require a separate \emph{GosipGUI} instance on each of these nodes. Of course it would be possible to extend the \emph{GosipGUI} software in a way that one central GUI communicates with several GOSIP IO servers at the PEXOR nodes via network protocols like HTTP or RPC. This would just require to change the GOSIP IO layer (see Fig. \ref{f-collaboration}) and extend the device selector in the generic GUI (Fig. \ref{f-commongui}) with a field to specify the PEXOR host. But still this was limited to display one front-end board at the same time only.

Furthermore, the idea of data servers for distributed process variables that can be monitored by remote GUIs is well established in existing frameworks like EPICS \cite{b-epics} or FESA \cite{b-fesa}, and needs not to be re-implemented in the scope of \emph{GosipGUI}. Instead, it is easily possible to develop EPICS IOC devices, or FESA plug-ins, to communicate with the front-ends via the GOSIP library. So large scale DAQ monitoring of GOSIP front-ends would be integrated into the experiment's standard control system. 

Besides of this,  the \emph{GosipGUI} can still have a qualification for large DAQ setups, as it provides an expert tool to check and investigate special front-end hardware. Such interactive control of all DAQ setup parameters may not be implemented to the standard control system which is rather intended to provide a monitoring of the overall state. So a \emph{GosipGUI} may be started by the DAQ expert for troubleshooting if the shift operator gets notified about a problem from the standard control system.

At the current state the \emph{GosipGUI} plays a good role for testing and commissioning of the DAQ hardware provided by the GSI Experiment Electronics department. This fills the niche between board development with configuration scripts and a full flavoured experiment control system. With this background, the \emph{GosipGUI} framework will be gradually improved and augmented with implementations for new front-end boards. 

The \emph{GosipGUI} software is freely avaliable under GPL \cite{b-subversion}.

\section*{Acknowledgment}
The authors thank Sven L\"ochner and Peter Wieczorek for valuable ideas and user feedback during the development and testing of  \emph{PolandGui} and \emph{ApfelGui}.

%
%
%
%
%
%

%


\end{document}